\documentclass[letter,aps,prl,twocolumn,superscriptaddress,showpacs,english,amsmaths,amssymb]{revtex4}

\usepackage{graphicx}
\usepackage{bm}

\begin{document}

\title{Pair correlations in a finite-temperature $1D$ Bose gas}

\author{K.V. Kheruntsyan}
\affiliation{ARC Centre of Excellence for Quantum-Atom Optics, Department of Physics, University of Queensland, Brisbane, Qld 4072, Australia}

\author{D.M. Gangardt}
\affiliation{Laboratoire Kastler-Brossel, Ecole Normale Sup\'erieure, 24 rue
Lhomond, 75005 Paris, France}

\author{P.D. Drummond}
\affiliation{ARC Centre of Excellence for Quantum-Atom Optics, Department of Physics, University of Queensland, Brisbane, Qld 4072, Australia}

\author{G.V. Shlyapnikov}
\affiliation{Laboratoire Kastler-Brossel, Ecole Normale Sup\'erieure, 24 rue
Lhomond, 75005 Paris, France}

\affiliation{\mbox{FOM Institute for Atomic and Molecular Physics, Kruislaan
407, 1098 SJ Amsterdam, The Netherlands}}

\affiliation{Russian Research Center Kurchatov Institute, Kurchatov
  Square, 123182 Moscow, Russia}

\date{\today}

\begin{abstract}
We calculate the two-particle local correlation for an interacting $1D$
Bose gas at finite temperature and classify various physical regimes. We present the
exact numerical solution by using the Yang-Yang equations and Hellmann-Feynman
theorem and develop analytical approaches. Our results draw prospects for
identifying the regimes of coherent output of an atom laser, and of finite-temperature 
``fermionization'' through the measurement of the rates of two-body inelastic
processes, such as photo-association. \end{abstract}

\pacs{05.30.-d, 03.75.Hh, 03.75.Pp, 05.30.Jp}

\maketitle

Recent observations of the one-dimensional ($1D$) regime in trapped
Bose gases \cite{Goerlitz2001-Schreck2001-Greiner2001} offer 
unique opportunities for extending our understanding of the physics of
these quantum many-body systems. The reason is that the $1D$ regime 
can be investigated theoretically by making use of the
known exactly solvable $1D$ models \cite{M}, which
have been the subject of extensive studies since the pioneering works
of Girardeau \cite{Girardeau1960}, Lieb and Liniger \cite{LiebLiniger1963},
and Yang and Yang \cite{Yang1969} (see \cite{Popov,Thacker1981,KorepinBook}
for reviews). With the development of experimentally viable Bose gases, 
these field theory models -- once of theoretical relevance only -- are now
becoming testable in table-top experiments.

The knowledge of the exact solutions to the $1D$ models allows us 
to go far beyond the mean-field Bogoliubov approximation. In the
current stage of studies of experimentally feasible $1D$ Bose gases, one of
the most important issues  that requires such an approach is understanding the
correlation properties in the various regimes at \emph{finite temperatures.} 

In this Letter we give an exact calculation of the finite-temperature two-particle 
local correlation for an interacting uniform $1D$ Bose gas, 
$g^{(2)}=\langle \hat{\Psi}^{\dagger }(x)^{2}\hat{\Psi}(x)^{2}\rangle /n^{2}$,
where $\hat{\Psi}(x)$ is the field operator and 
$n=\langle \hat{\Psi}^{\dagger }(x)\hat{\Psi}(x)\rangle$
is the linear ($1D$) density. 
As a result, we identify 
and classify various finite-temperature regimes of the 1D Bose gas.
Aside from this, the pair correlations are responsible for 
the rates of inelastic collisional processes \cite{Kagan1985}, 
and are of particular importance for the studies of coherence properties 
of atom ``lasers'' produced in $1D$ waveguides.       

At $T=0$, the local two- and three-particle correlations of a uniform 
$1D$ Bose gas have recently been calculated in Ref. \cite{Gangardt}. 
Here one has two well-known and physically distinct
regimes of quantum degeneracy. For weak couplings or high densities,
the gas is in a coherent or Gross-Pitaevskii (GP) regime with 
$g^{(2)}\rightarrow 1$. In this regime, long-range order is destroyed by 
long-wavelength phase fluctuations \cite{Haldane} and the equilibrium 
state is a quasi-condensate characterized by suppressed density fluctuations.
For strong couplings or low densities, 
the gas reaches the strongly interacting or Tonks-Girardeau (TG) regime 
and undergoes ``fermionization'' \cite{Girardeau1960,LiebLiniger1963}: 
the wave function strongly decreases as particles approach 
each other, and $g^{(2)}\rightarrow 0$.

At the non-zero temperatures studied here, one has to further extend the 
classification of different regimes. For strong enough couplings or low densities, 
we obtain the TG regime with $g^{(2)}\rightarrow 0$ not only at low 
temperatures, but also at high temperatures. In addition, and in contrast 
to previous $T=0$ results, we find a weak-coupling or high-density
regime in which fluctuations are enhanced. Asymptotically, they reach
the non-interacting Bose gas level of $g^{(2)}\rightarrow 2$ (rather
than $g^{(2)}\rightarrow 1$), for any finite temperature $T$.

The emergence of this behavior at low temperatures implies that one
can identify \emph{three} physically distinct regimes of quantum degeneracy:
the strong-coupling TG regime of ``fermionization'' with $g^{(2)}\rightarrow 0$,
a coherent GP regime with $g^{(2)}\simeq 1$ at intermediate coupling
strength, and a fully decoherent quantum (DQ) regime with $g^{(2)}\simeq 2$
at very weak couplings. In the GP regime, where the density
fluctuations are suppressed and one has a quasi-condensate, the local correlation
approaches the coherent level of $g^{(2)}\simeq 1$.
However, a free (non-interacting) Bose gas at any finite $T$ must have
$g^{(2)}=2$. So, below a critical density- and temperature-dependent level of
interaction strength one must have an increase in thermal fluctuations, until
the ideal-gas level is reached in a continuous transition. At $T=0$
the transition is discontinuous and occurs at zero interaction strength,
so that it can be viewed as a zero-temperature phase transition. At
high temperatures the GP regime vanishes and a decrease of coupling
strength transforms the high-temperature TG regime directly into a classical 
ideal gas.

We start by considering a uniform gas of $N$ bosons interacting via repulsive delta-function
potential in a $1D$ box of length $L$ with periodic boundary condition.
In the thermodynamic limit ($N,L \rightarrow \infty$, while the density $n=N/L$ 
is fixed) the system is exactly integrable by using the Bethe ansatz, both at
$T=0$ and finite temperature \cite{LiebLiniger1963,Yang1969}. 
In second quantization, the Hamiltonian is
\begin{equation}
\hat{H}=\frac{\hbar ^{2}}{2m}\int dx\, \partial _{x}
\hat{\Psi}^{\dagger }\partial _{x}\hat{\Psi} +\frac{g}{2}\int dx\, \hat{\Psi}^{\dagger }\hat{\Psi}^{\dagger }\hat{\Psi} \hat{\Psi},
\end{equation}
where $\hat{\Psi}(x)$ is the field
operator, $m$ is the mass, and $g>0$ is the
coupling constant. For Bose gases in highly elongated traps to be described by this $1D$ model, the coupling $g$ is expressed 
through the $3D$ scattering length $a$. This is done assuming that $a$ is much smaller than the amplitude of transverse
zero point oscillations $l_{0}=\sqrt{\hbar/m\omega_0}$, where $\omega_0$ is
the frequency of the transverse harmonic potential. For a positive 
$a\ll l_{0}$ one has 
\begin{equation}
g=2\hbar ^{2}a/ml_{0}^{2},\label{g}
\end{equation}
and the $1D$ scattering length $a_{1D}$ is $\hbar ^{2}/mg\simeq l_{0}^{2}/a\gg l_{0}$ \cite{Olshanii98}. The
$1D$ regime is reached if $l_{0}$ is much smaller than the thermal de Broglie wavelength
of excitations $\Lambda _{T}=(2\pi \hbar ^{2}/mT)^{1/2}$ and a characteristic length scale $l_{c}$ \cite{lc} responsible for short-range correlations. On
the same grounds as at $T=0$ \cite{Gangardt}, one finds that for fulfilling
this requirement it is sufficient to satisfy the inequalities $a\ll l_{0}\ll
\{1/n,\, \Lambda _{T}\}$. 

The uniform $1D$ Bose gas with a short-range repulsive
interaction can effectively be characterized by two parameters:
the dimensionless coupling parameter 
\begin{equation}
\gamma =mg/\hbar ^{2}n,
\end{equation}
and the reduced temperature $\tau =T/T_{d}$, where the temperature
of quantum degeneracy is given by $T_{d}=\hbar ^{2}n^{2}/2m$, in
energy units ($k_{B}=1$).

For calculating the local two-body correlation $g^{(2)}$
at any values of $\gamma$ and $\tau$ we use the Hellmann-Feynman
theorem \cite{Hellmann1933-Feynman1939}.
At $T=0$, it has been used for calculating the mean interaction energy
\cite{LiebLiniger1963}, and for expounding the issue of local pair
correlations \cite{Gangardt}. Consider the partition function $Z=\exp
(-F/T)=\mathrm{Tr}\exp(-\hat{H}/T)$ which determines the free energy $F$. Here
the trace is taken over the states of the system with a fixed number of
particles in the canonical formalism. For the grand canonical description, one
has to replace the condition of a constant particle number by the condition of
a constant chemical potential $\mu$ and add a term $-\mu\hat{N}$ to the
Hamiltonian. For the derivative of $F$ with respect to the
coupling constant one has $\partial F/\partial g=-T\partial(\log Z)/\partial
g$ and hence
\begin{equation}
\frac{\partial F}{\partial g}=\frac{1}{Z}{\text{Tr}}\left[  e^{-\hat{H}/T}
\;\partial\hat{H}/\partial g\right]  =(L/2)\langle\hat{\Psi}^{\dagger}%
\hat{\Psi}^{\dagger}\hat{\Psi}\hat{\Psi}\rangle\,\,.\label{eq:hellfeyn}
\end{equation}
Introducing the free energy per particle $f(\gamma ,\tau )=F/N$,
the normalized two-particle correlation is:
\begin{equation}
g^{(2)}=\frac{ \langle \hat{\Psi}^{\dagger }\hat{\Psi} ^{\dagger }
\hat{\Psi} \hat{\Psi} \rangle }{n^{2}}=\frac{2m}{\hbar ^{2}n^{2}}\left(\frac{\partial
f(\gamma ,\tau )}{\partial \gamma }\right)_{n,\tau }\,
.\label{gnorm}
\end{equation}

\begin{figure}
\includegraphics[width=8cm]{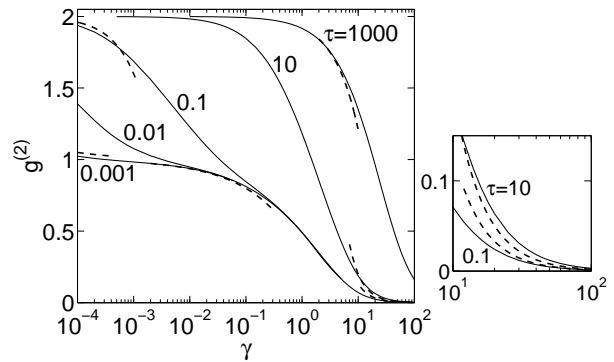}
\caption{The local correlation $g^{(2)}$ versus $\gamma $ at different $\tau $.
The solid curves are exact numerical results, while the dashed curves represent
analytic results (see text).}
\label{g2vsgamma_log}
\end{figure}

We have calculated the free energy $f(\gamma ,\tau )$ by numerically
solving the Yang-Yang exact integral equations for the excitation
spectrum and the distribution function of ``quasi-momenta'' \cite{Yang1969}.
By implementing a post-selective algorithm that ensures that the derivative
of $f(\gamma ,\tau )$ is taken for constant $n$, we then calculate
$g^{(2)}$ from Eq. (\ref{gnorm}). The results of our calculations
are presented in Fig.~\ref{g2vsgamma_log}.
We now give a physical description of different regimes determined
by the values of the coupling constant $\gamma $ and the reduced
temperature $\tau $.

\textbf{\textit{Strong coupling regime}} [$\gamma \gtrsim \max (1,\sqrt{\tau })$].
In the strong coupling TG regime the local correlation $g^{(2)}$
reduces dramatically due to the strong repulsion between particles,
and becomes zero for $\gamma\rightarrow\infty$. In this regime the physics
resembles that of free fermions, both below and above the quantum
degeneracy temperature. Along the lines of Ref. \cite{Gangardt},
to leading order in $1/\gamma $ the finite-temperature $g^{(2)}$
can be expressed through derivatives of Green's function of free fermions
$G(x)=\int dkn_{F}(k)\exp {(ikx)}/(2\pi )$, where $n_{F}(k)$ are
occupation numbers for free fermions. For the normalized local correlation
we obtain 
$g^{(2)}=4\left[\left(G^{\prime }(0)\right)^{2}-G^{\prime \prime }(0)G(0)\right]/\gamma ^{2}n^{4}$.

In the regime of quantum degeneracy, $\tau \ll 1$, the local correlation
is dominated by the ground state distribution $n_{F}(k)=\theta (k_{F}^{2}-k^{2})$,
where $k_{F}=\pi n$ is the Fermi momentum. Small finite-temperature
corrections are obtained using the Sommerfeld expansion:
\begin{equation}
g^{(2)}=\frac{4}{3}\left(\frac{\pi }{\gamma }\right)^{2}\left[1+\frac{\tau ^{2}}{4\pi ^{2}}\right],\; \; \tau \ll 1\, \, \, .\label{eq:II}
\end{equation}
This low temperature result for $g^{(2)}$ has a simple physical
meaning. A characteristic distance related to the interaction between
particles is $a_{1D}=\hbar ^{2}/mg\sim 1/\gamma n$, and fermionic
correlations are present at interparticle distances $x\gtrsim a_{1D}$.
For smaller $x$ the correlations practically do not change.
Hence, the correlation $g^{(2)}$ for large
$\gamma $ is nothing else than the pair correlation for
free fermions at a distance $a_{1D}$. 
The latter is $g^{(2)}\sim (k_{F}a_{1D})^{2}\sim 1/\gamma ^{2}$,
which agrees with the result of Eq.(\ref{eq:II}) for $\tau \rightarrow 0$.

In the temperature interval $1\ll \tau \ll \gamma ^{2}$ the gas is
nondegenerate, but the $1D$ scattering length $a_{1D}$ is still much smaller
than the thermal de Broglie wavelength $\Lambda _{T}$. As the characteristic
momentum of particles is now the thermal momentum $k_{T}=\sqrt{2mT}/\hbar $,
one estimates that $g^{(2)}\sim (k_{T}a_{1D})^{2}\sim \tau /\gamma ^{2}$.
Calculating Green's function $G(x)$ for the classical distribution
$n_{F}(k)$, we obtain:
\begin{equation}
g^{(2)}=2\tau /\gamma ^{2},\; \; 1\ll \tau \ll \gamma ^{2},\label{eq:III}
\end{equation}
which agrees with the given qualitative estimate. The local correlation $g^{(2)}$ 
is still much smaller than one and we thus have
a regime of high-temperature ``fermionization''.

The results of Eqs. (\ref{eq:II}) and (\ref{eq:III}) agree with
the outcome of our numerical calculations. For $\tau =0.1$ and $\tau =10$,
they are shown in Fig.~\ref{g2vsgamma_log} (in the region of large
$\gamma $) by dashed curves next to the solid curves found numerically
for the same values of $\tau $.

\textbf{\textit{GP regime}} (\textit{$\tau ^{2}\lesssim \gamma \lesssim 1$}).
In the intermediate coupling or GP regime, for sufficiently low temperatures
the equilibrium state is a quasi-condensate: the density fluctuations
are suppressed, but the phase fluctuates \cite{Mermin1966}. As the
phase coherence length $l_{\phi }$ greatly exceeds the short-range characteristic
length $l_{c}=\hbar /\sqrt{mng}$, for finding the local correlations
the field operator can be represented as a sum of the macroscopic
component $\Psi _{0}$ and a small component $\hat{\Psi}^{\prime }$ describing
finite-momentum excitations. Actually, the component $\Psi _{0}$
contains the contribution of excitations with momenta $k\lesssim k_{0}\ll l_{c}^{-1}$,
whereas $\hat{\Psi}^{\prime }$ includes the contribution of larger $k$.
At the same time, the momentum $k_{0}$ is chosen such that most of
the particles are contained in $\Psi _{0}$. This picture
is along the lines of Ref. \cite{Popov}, and the momentum $k_{0}$
drops out of the answer as the main contribution of $\hat{\Psi}^{\prime }$ 
to $g^{(2)}$ is provided
by excitations with $k\sim l_{c}^{-1}$ \cite{comment1}. The two-particle
local correlation is then reduced to $g^{(2)}=1+2(\langle \hat{\Psi}^{\prime\dagger}\hat{\Psi}^{\prime}\rangle +\langle \hat{\Psi}^{\prime } \hat{\Psi}^{\prime }\rangle )/n$.
The normal and anomalous averages, $\langle \hat{\Psi} ^{\prime\dagger }\hat{\Psi}^{\prime }\rangle$ and $\langle \hat{\Psi}^{\prime }\hat{\Psi}^{\prime }\rangle$, 
can be calculated by using the same Bogoliubov transformation for 
$\hat{\Psi}^{\prime }$ as in 3D. This gives the result that 
\begin{equation}
g^{(2)}=1+\int _{-\infty }^{\infty }\frac{dk}{2\pi n}\left[\frac{E_{k}}{\varepsilon _{k}}(1+n_{k})-1\right]\, ,\label{eq:g2weak}
\end{equation}
where $E_{k}=\hbar ^{2}k^{2}/2m$, $\varepsilon _{k}=\sqrt{E_{k}^{2}+2ngE_{k}}$
is the Bogoliubov excitation energy, and $n_{k}$ are occupation numbers
for the excitations.

The integral term in Eq. (\ref{eq:g2weak}) contains the contribution
of both vacuum and thermal fluctuations. The former is determined
by excitations with $k\sim l_{c}^{-1}$, and at $\tau =0$ we immediately
recover the zero-temperature result of Ref. \cite{Gangardt}. For
very low temperatures $\tau \ll \gamma $, thermal fluctuations give
an additional correction, so that 
\begin{equation}
g^{(2)}=1-2\sqrt{\gamma }/\pi +\pi \tau ^{2}/(24\gamma ^{3/2}),\; \; \tau \ll \gamma \ll 1\, .\end{equation}
The phase coherence length is determined by vacuum fluctuations of
the phase and is $l_{\phi }\sim l_{c}\exp {(\pi /\sqrt{\gamma })}$
\cite{Haldane}. For $\tau =0.001$ the above approximate result,
shown in Fig.~\ref{g2vsgamma_log} at intermediate values of $\gamma $,
practically coincides with the corresponding exact numerical result.

For temperatures $\tau \gg \gamma $, thermal fluctuations are more
important than vacuum fluctuations. The main contribution to the local
correlation is again provided by excitations with $k\sim l_{c}^{-1}$,
and we obtain, from Eq. (\ref{eq:g2weak}) 
\begin{equation}
g^{(2)}=1+\tau /(2\sqrt{\gamma })\, ,\; \; \gamma \ll \tau \ll \sqrt{\gamma }.\label{eq:V}
\end{equation}
The phase coherence length is determined by long-wavelength phase
fluctuations. The calculation similar to that for a trapped gas 
\cite{Petrov} gives $l_{\phi }\approx \hbar ^{2}n/mT$. The
condition $l_{\phi }\gg l_{c}$, which is necessary for the existence
of a quasi-condensate and for the applicability of the Bogoliubov
approach, immediately yields the inequality $\tau \ll \sqrt{\gamma }$.
Thus, Eq. (\ref{eq:V}) is valid under the condition $\gamma \ll \tau \ll \sqrt{\gamma }$,
and the second term in the rhs of this equation is a small correction.
One can easily see that this correction is just the relative mean
square density fluctuations. In the region of its validity, the result
of Eq.~(\ref{eq:V}) agrees well with our numerical data, and is
shown in Fig.~\ref{g2vsgamma_log} for $\tau =0.001$,
in the region of small $\gamma $ values. The exact results graphed
in Fig.~\ref{g2vsgamma_log} for different values of $\tau$ show
that the coherent or GP regime is not present for $\tau \gtrsim 0.1$,
in the sense that $g^{(2)}$ as a function of $\gamma $ does not
develop a plateau around the value $g^{(2)}=1$.

\textbf{\emph{Decoherent regime:}} At very weak couplings, $\gamma \lesssim min(\tau ^{2},\sqrt{\tau })$, the gas enters a decoherent
regime. Both phase and density fluctuations are large.
At small enough $\gamma$ the local correlation is always
close to the result for free bosons, $g^{(2)}=2$. In this regime, the only consequence of quantum degeneracy
is the quantum Bose distribution for occupation numbers of particles,
so we can further divide it into a decoherent quantum (DQ) regime
for $\tau <1$ and a decoherent classical (DC) regime for $\tau >1$.

The result of Eq. (\ref{eq:V}) cannot be used for $\gamma =0$. In
this case one has a gas of free bosons, and Wick's theorem leads to
$g^{(2)}=2$ at any $\tau $. For small $\tau $, our data
in Fig. \ref{g2vsgamma_log} show a sharp increase of $g^{(2)}$
from almost $1$ to almost $2$ when $\gamma\lesssim\tau ^{2}$.
This is a continuous transition from the quasi-condensate to the DQ
regime \cite{Castin}. Lowering the temperature lowers the value of $\gamma $ at
which this transition occurs. For $\tau =0$ the transition takes
place at $\gamma =0$. In this case it is discontinuous and can be
regarded as a zero-temperature phase transition.

The DQ regime can be treated asymptotically by employing a standard
perturbation theory with regard to the coupling constant $g$. Omitting
the details of calculations, which will be published elsewhere, for
the local correlation we obtain
\begin{equation}
g^{(2)}=2-4\gamma /\tau ^{2},\, \, \, \, \, \sqrt{\gamma }\ll \tau \ll 1.
\label{eq:g2lowT}
\end{equation}

At higher temperatures $\tau \gtrsim 1$ the decoherent \emph{quantum}
regime ($\sqrt{\gamma }\lesssim \tau \lesssim 1$) transforms to the
decoherent \emph{classical} regime ($\tau \gtrsim {\textrm{max}}\{1,\gamma ^{2}\}$)
and $g^{(2)}$ remains close to $2$ (see Fig.~\ref{g2vsgamma_log}).
The local correlation is found in the same way as in the
DQ regime and takes the asymptotic form 
\begin{equation}
g^{(2)}=2-\gamma \sqrt{2\pi /\tau },\; \; \tau \gg {\textrm{max}}\{1,\gamma ^{2}\}.\label{eq:g2perthighT}\end{equation}

The result of Eq. (\ref{eq:g2perthighT}) remains valid for large values
of $\gamma $, provided that $\gamma ^{2}\ll \tau$. Here the de
Broglie wavelength $\Lambda _{T}$ becomes smaller than $a_{1D}$ and
the regime of high-temperature ``fermionization'' continuously
transforms into the decoherent regime of a classical gas. The corrections
to $g^{(2)}=2$, given by Eqs. (\ref{eq:g2lowT}) and (\ref{eq:g2perthighT}),
are in agreement with the exact numerical calculations. In Fig.~\ref{g2vsgamma_log},
the approximate analytical results are shown by the dashed lines next to the corresponding solid lines found numerically, for $\tau =0.1$
and $\tau =1000$.

In conclusion, we have calculated the two-particle local correlation $g^{(2)}$
for a uniform $1D$ Bose gas at finite temperatures \cite{comment2}. Within
their range of validity, the analytical results agree with exact numerical
calculations based on the Hellmann-Feynman theorem and the Yang-Yang
equations. The knowledge of $g^{(2)}$ allows one to deduce when the
approximate GP equation often used for first-order phase-coherence
calculations is valid. The value of $g^{(2)}\simeq1$ indicates that the
correlation function factorizes, which is a necessary condition for using the
GP approach. The prediction of coherent behavior only for certain temperatures
and interaction strengths, $\tau^{2}\lesssim\gamma\lesssim1$\textit{,} may be
an important criterion for atom lasers where spatial coherence is a necessary
ingredient in obtaining interference and high-resolution interferometry. Our
results are also promising for identifying the regime of ``fermionization'' in
finite-temperature $1D$ Bose gases through the measurement of inelastic
processes in pair interatomic collisions such as photo-association. Here the
transitions occur at interatomic distances which are much smaller than the
short-range correlation length $l_{c}$ and, therefore, the rate will be
proportional to the local correlation $g^{(2)}$.

We also find a fully decoherent quantum regime in the case of very weak
interactions or high densities. There is a \emph{continuous} transition from the GP regime to the ideal gas limit ($\gamma\rightarrow 0$) where the gas displays large thermal (Gaussian) density fluctuations with $g^{(2)}=2$ at any finite temperature \cite{ZiffUhlenbeckKac77}. As $\gamma$ is decreased towards the ideal gas, the GP result only holds above a certain ratio of interaction strength to density. Below this, the GP approach becomes invalid and there is a dramatic increase in fluctuations, with $g^{(2)}\rightarrow 2$ as $\gamma\rightarrow 0$.

\begin{acknowledgments}
K.K. and P.D. acknowledge the ARC and the Humboldt foundation for
their support. D.G. and G.S. acknowledge support from the French Minist\`{e}re
des Affaires Etrang\`{e}res, from the Dutch Foundations NWO and FOM,
from INTAS, and from the Russian Foundation for Basic Research. We acknowledge M. Cazalilla for drawing our attention to a typographical 
error 
in Eq. (6) of the original version of the manuscript. Laboratoire Kastler Brossel is a research unit of Universit\'{e} Pierre et Marie Curie and ENS, associated with CNRS (UMR 8552).
\end{acknowledgments}


\begin{thebibliography}{10}
\bibitem{lc} Well below the temperature of quantum degeneracy, $l_c=\hbar/\sqrt{m\mu}$ in the GP and TG regimes ($\mu$ is the
chemical potential), and otherwise $l_c$ is of the order of $\Lambda _{T}$. In the GP regime $l_c$
coincides with the healing length, $l_c=\hbar/\sqrt{mng}$, and in the TG regime $l_c\sim 1/n$.
\bibitem{comment1}For $\tau \ll \gamma $, the contribution of
thermal fluctuations to $g^{(2)}$ is not important and is provided by
excitations with $k\sim (\tau /l_{c}\gamma )$. The momentum $k_{0}$ can always
be chosen smaller than these values of $k$.
\end{thebibliography}
\end{document}